\documentclass[aps,prr,longbibliography,superscriptaddress,twocolumn,showkeys,amsmath,amssymb,floatfix]{revtex4-2}

\usepackage[english]{babel}
\usepackage{amsmath,amssymb,bbm,mathrsfs,bm,braket,color,graphicx,comment,amsfonts,dsfont}
\usepackage{siunitx,physics}
\usepackage[colorlinks,citecolor=blue,urlcolor=blue]{hyperref}
\usepackage[mathscr]{euscript}
\usepackage{import}

\newcommand\fig[3]{\begin{figure}\centering\includegraphics[#2]{fig_#1}\caption{\label{fig:#1} #3}\end{figure}}

\begin{document}
\title{Chirality flip of Weyl nodes and its manifestation in strained MoTe$_2$}

\author{Viktor K\"{o}nye}
\affiliation{Institute for Theoretical Solid State Physics, IFW Dresden and W\"urzburg-Dresden Cluster of Excellence ct.qmat, Helmholtzstr. 20, 01069 Dresden, Germany}

\author{Adrien Bouhon}
\affiliation{Nordic Institute for Theoretical Physics (NORDITA), Stockholm, Sweden}

\author{Ion Cosma Fulga}
\affiliation{Institute for Theoretical Solid State Physics, IFW Dresden and W\"urzburg-Dresden Cluster of Excellence ct.qmat, Helmholtzstr. 20, 01069 Dresden, Germany}

\author{Robert-Jan Slager}
\affiliation{TCM Group, Cavendish Laboratory, University of Cambridge, J. J. Thomson Avenue, Cambridge CB3 0HE, United
Kingdom}

\author{Jeroen van den Brink}
\affiliation{Institute for Theoretical Solid State Physics, IFW Dresden and W\"urzburg-Dresden Cluster of Excellence ct.qmat, Helmholtzstr. 20, 01069 Dresden, Germany}
\affiliation{Institute  for  Theoretical  Physics,  TU  Dresden,  01069  Dresden,  Germany}

\author{Jorge I. Facio}
\affiliation{Institute for Theoretical Solid State Physics, IFW Dresden and W\"urzburg-Dresden Cluster of Excellence ct.qmat, Helmholtzstr. 20, 01069 Dresden, Germany}

\date{\today}

\begin{abstract}
Due to their topological charge, or chirality, the Weyl cones present in topological semimetals are considered robust against arbitrary perturbations.
One well-understood exception to this robustness is the pairwise creation or annihilation of Weyl cones, which involves the overlap in energy and momentum of two oppositely charged nodes.
Here we show that the topological charge can in fact change sign, in a process that involves the merging of not two, but three Weyl nodes.
This is facilitated by the presence of rotation and time-reversal symmetries, which constrain the relative positions of Weyl cones in momentum space.
We analyze the chirality flip process, showing that transport properties distinguish it from the conventional, double Weyl merging.
Moreover, we predict that the chirality flip occurs in MoTe$_2$, where experimentally accessible strain leads to the merging of three Weyl cones close to the Fermi level.
Our work sets the stage to further investigate and observe such chirality flipping processes in different topological materials.
\end{abstract}
\maketitle

\section{Introduction} 
Weyl semimetals are arguably the most robust form of gapless topological matter \cite{Nielsen1981,Murakami2007,Wan2011,Burkov2011a,Burkov2011,Chiu2016,Yan2017,Armitage2018,Bernevig2018}.
They host point-like degeneracies of their energy bands, called Weyl nodes (or cones), whose low energy, linear dispersion relation is similar to that of elementary particles called Weyl fermions \cite{Weyl1929}.
These band touching points are not accidental, but are a manifestation of the topologically nontrivial character of the semimetal phase.
Each node has a chirality associated to it, a topological charge given by its Chern number, which means that an isolated Weyl cone cannot be gapped out by any perturbation.
Instead, it can only be moved in energy and momentum space, or tilted \cite{Soluyanov2015}.
This property sets Weyl semimetals apart from the many other types of gapless topological matter, in which degenerate points \cite{Young2012, Wang2012, Bradlyn2016, Lepori2016, Lepori2016b, Fulga2017} (or lines \cite{Burkov2011, Fang2016}) require additional symmetries in order to remain protected.

Provided that Weyl cones are not isolated, it is possible to change their number: 
they can be created or annihilated pairwise when they overlap in energy and momentum.
Similar to electrodynamics, these two-node processes require `topological charge conservation,' meaning that the total Chern number of all band touching points in the Brillouin zone (BZ) must vanish \cite{Nielsen1981}.
The pairwise creation and annihilation of Weyl cones has been studied extensively, especially as a method to engineer Weyl semimetal phases by applying perturbations such as strain \cite{Sie2019,PhysRevB.90.155316,weylbbctop,PhysRevLett.121.246403}, magnetic field or changes in magnetization direction \cite{Zhang2017, Cano2017, PhysRevResearch.1.032044,ray2020tunable}, disorder \cite{BbcWeyl, phasediagweyl, pixley2016rare, pixley2021rare}, phonons \cite{Singh2020}, or high-frequency illumination \cite{Fu2017, Fu2019} to real materials.

In theory, there is no constraint limiting the number of Weyl nodes that can merge at a given point in the BZ.
In practice, however, processes involving the simultaneous overlap of three or more cones of different chirality are highly improbable, especially when each node is allowed to occupy any point in energy and momentum space.
To date, multi-Weyl merging has been studied mainly in the context of rather exotic types of topological semimetals \cite{Fang2012, Bradlyn2016}, hosting `unconventional fermions.' There, the symmetry-protected band degeneracies may split into multiple Weyl nodes if very particular symmetries are broken \cite{Bradlyn2016, Fulga2018, Thirupathaiah2021}.

Here we study simple three-node processes occurring purely inside a Weyl semimetal phase, without unconventional fermions: a single Weyl cone overlaps with a pair of oppositely charged ones, causing the latter to disappear from the band structure.
By topological charge conservation, the result is a single node with a flipped chirality.
Interestingly, while lattice symmetries are not responsible for the existence of the degeneracy points, they play a fundamental role in enabling the three-node process to occur.
This is because they restrict the relative positions and charges of the nodes in the BZ (for instance to planes or lines).
Throughout the following, we will focus on systems obeying a combination of twofold rotation and time-reversal symmetry.

One of the advantages of studying multi-node processes in Weyl semimetals as opposed to more exotic types of gapless topological matter is that Weyl semimetal phases are much more abundant in real materials \cite{Xu2020}.
In fact, we find that one of the earliest predicted Weyl materials, MoTe$_2$ \cite{PhysRevX.6.031021}, hosts very close to the Fermi level a three-node process upon applying small uniaxial strain. As a result of this process, a Weyl node in a high-symmetry plane flips its chirality.
This shows that the three-node process is within reach of photoemission experiments.
Further, it shows that strain in real materials can act as a `chirality switch', allowing to tune the transport properties associated with gapless topological systems.

\section{Chirality flip process in simple models} 
We start by discussing how the chirality of a Weyl node can flip using toy models. 
Given the requirement of topological charge conservation and considering only simple Weyl cones, chirality flips are only possible if at least two other, oppositely charged Weyl nodes are involved.
For concreteness, we consider processes that start from three Weyl nodes with Chern numbers ${\cal C}=+1$, $-1$, $-1$, and end with a single Weyl node having ${\cal C}=-1$ \footnote{We calculate the Chern number using the occupied bands with the Berry curvature defined as $\vb{B}_{n,k}=\grad_k\cross\vb{A}_{n,k}$, where $\vb{A}_{n,k}=-i\bra{n,k}\grad_k\ket{n,k}$. This definition is chosen to be consistent with the FPLO code.}. We find that there are two scenarios through which this can happen: one that involves a three-node process and one that only involves two-node processes.

We begin by discussing the first scenario, corresponding to a three-node process, meaning that all three Weyl cones simultaneously overlap at the same point in the BZ. 
While this merging could in principle occur at generic momenta, its likelihood can be greatly increased when symmetries constrain the relative positions and charges of the band touching points. 
As mentioned above, we will consider a continuum model invariant under the combination of twofold rotation and time-reversal symmetries, with Hamiltonian
\begin{equation}
\label{eq:simpleH}
    H_1(\vb{k}) = k_x \sigma_x + \left(\alpha k_z + k_z^3 \right) \sigma_y + k_y \sigma_z.
\end{equation}
Here, $\vb{k}=(k_x, k_y, k_z)$ is the quasimomentum and $\sigma_i$ are Pauli matrices encoding the degree of freedom associated with the two bands.
For simplicity, we will set units such that the Hamiltonian and the quasimomentum are dimensionless. 

The Hamiltonian Eq.~\eqref{eq:simpleH} obeys a twofold rotation symmetry along the $k_x=k_y=0$ axis, with operator $C_{2}=-i\sigma_y$, as well as time-reversal symmetry, $T=i\sigma_y K$ (where $K$ is complex conjugation), such that
\begin{align}
    TH^{\phantom\dag}_1(\vb{k})T^\dag &= H^{\phantom\dag}_1(-\vb{k}),\\ 
    C^{\phantom\dag}_{2}H^{\phantom\dag}_1(\vb{k})C_{2}^\dag &= H^{\phantom\dag}_1(-k_x,-k_y,k_z).
\end{align}
Their combination, $C_{2}T=K$ implies that
\begin{equation}\label{eq:c2t}
    H_1^*(k_x,k_y,k_z) = H_1^{\phantom\dag}(k_x,k_y,-k_z),
\end{equation}
and that the Hamiltonian is real for $k_z=0$. 

The constraint Eq.~\eqref{eq:c2t} means that if a Weyl cone is present in the $C_{2}T$ invariant plane, $k_z=0$, then it cannot exit the plane, due to topological charge conservation. Further, if there is a Weyl node at $k_z>0$, another one must be positioned symmetrically at $k_z<0$. Note that pairs of out-of-plane cones must have the same Chern number, since both time-reversal and twofold rotation are charge preserving.

For $\alpha<0$, $H_1$ hosts three Weyl cones, a positive chirality node at the $\Gamma$ point, $\vb{k}=0$, and two negative chirality nodes at $\vb{k}=(0,0,\pm\sqrt{|\alpha|})$. 
In contrast, for $\alpha>0$ there is a single Weyl node with negative chirality at $\Gamma$. 
As shown in Fig.~\ref{fig:simple_disprel}(a, b), changing the parameter $\alpha$ continuously from negative to positive values causes the three nodes to merge at the same point, such that the central node flips its chirality.

The three-node process occurring in the band structure of $H_1$ is protected by symmetry: the Weyl merging \emph{must} involve all three nodes simultaneously. 
This is because the Hamiltonian obeys $C_{2}$ and $T$ separately, and because all Weyl nodes are located on the twofold rotation axis.
The band touching points cannot move away from $k_{x,y}=0$ without breaking the rotation symmetry, and they must be positioned symmetrically with respect to $k_z$ due to time-reversal symmetry. 
The only allowed overlap is thus at $\Gamma$. 

The presence of a chirality flip in $H_1$ is also related to topological quantities. 
Due to the presence of $C_2T$, the Hamiltonian is real at $k_z=0$. 
Therefore, a closed momentum-space loop in this plane will be characterized by a $\pi$-quantized Berry phase, provided that the loop encircles the central node. 
This $\mathbb{Z}_2$ topological invariant indicates that the parity of nodes inside the loop must be conserved \cite{bouhon2019wilson, magtop}. Therefore, the central node is not allowed to leave the plane, but it is allowed to reverse its chirality.

\fig{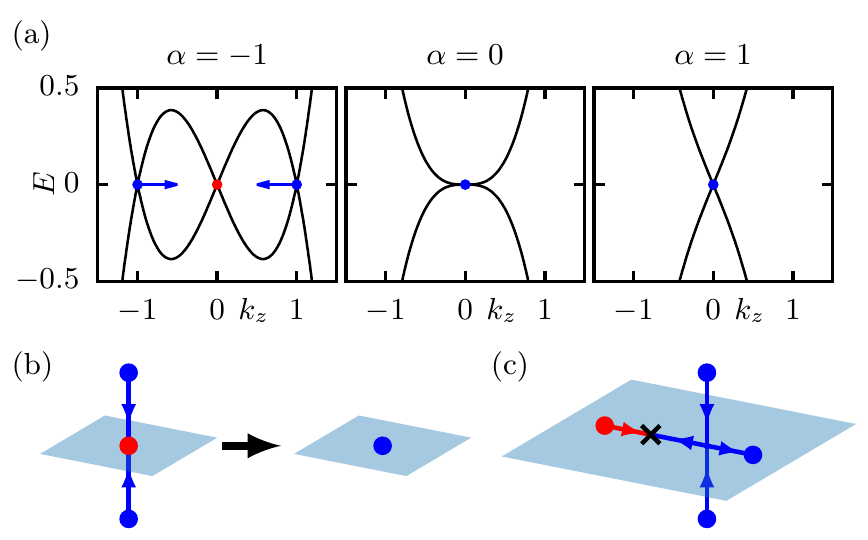}{width=8.6cm}{Schematic representation of the Weyl node processes. The red (blue) points represent positive (negative) chirality Weyl nodes, the arrows represent their trajectories upon increasing $\alpha$, and the $C_2T$-invariant plane is marked in blue. (a) Band structure of Eq.~\eqref{eq:simpleH} plotted at $k_x=k_y=0$ for three values of $\alpha$. (b) Chirality flip with a three-node process, as occurring in $H_1$. (c) Chirality flip with two-node processes, as happens in the Hamiltonian $H_2$.}

If the system instead hosts Weyl cones away from the rotation axis, or if it obeys only the combined $C_2T$ symmetry but not its individual components, the three-node process is not symmetry protected anymore.
This does not mean that it is forbidden.
Even if it is not forced to occur, it still can, and as we show later actually does happen in real materials.

We now move on to discuss the second scenario for the chirality flip, one that does not involve a three-node process. We illustrate this scenario in Fig.~\ref{fig:simple_disprel}(c), which shows the Weyl points of the Hamiltonian
\begin{equation}
\label{eq:simpleHboring}
    H_2(\vb{k}) = k_y\sigma_x + k_x k_z\sigma_y + \left(k_x^2+k_x^3-k_z^2-\alpha\right)\sigma_z.
\end{equation}
Note that $H_2$ obeys the same constraint as $H_1$ in Eq.~\eqref{eq:c2t}, but has neither rotation nor time-reversal symmetries. 

For $\alpha<0$ there are three Weyl nodes in the band structure of $H_2$: 
one in the $C_2T$ invariant plane $\vb{k_c}=(k_{0},0,0)$, where $k_{0}$ is the single real solution of $k^3+k^2=-|\alpha|$, and two away from the $C_2T$ plane at $\vb{k_s}=(0,0,\pm\sqrt{|\alpha|})$. 
Due to $C_2T$, the outside nodes can only enter the plane at the same point, as before.
However, now they can meet without the third node, thus forming a double Weyl point.
In $H_2$, the double Weyl point occurs at $k_z=0$ when $\alpha=0$.
For $\alpha\gtrapprox 0$ the overlapping nodes separate again, and all three cones are in the $C_2T$ plane, located at $(k_{i},0,0)$, where $k_{i}$ are the real solutions of $k^3+k^2=|\alpha|$. 
The three nodes remain until $\alpha=4/27$, at which point two of the opposite chirality Weyl points annihilate. At $\alpha>4/27$, we are left with a single Weyl node. 
The trajectory of all three Weyl cones is shown schematically in Fig.~\ref{fig:simple_disprel}(c).
Similar to the behavior of nodes in the spectrum of $H_1$, going from negative to positive $\alpha$ converts a positive chirality Weyl point into a negative chirality one.

As before, considering a momentum-space loop in the $C_2T$ plane that encloses the projections of all nodes in Fig.~\ref{fig:simple_disprel}(c), we find a $\pi$-quantized Berry phase for all $\alpha$. 
This indicates that the two-node and three-node process cannot be distinguished topologically by using the Berry phase. 
This is expected, as one can imagine continuously moving the meeting point of the out-of-plane nodes until this point overlaps with the $k_z=0$ node. However, when considering an infinitesimally small loop around the initial red node, a blue node will have to cross this loop for some $\alpha$, thus ensuring a discontinuity in the Berry phase. 

We note that $C_2T$-symmetry also relates to the Euler class \cite{Eulerdrive, BJY_nielsen, jiang2021observation, bouhon2020geometric,Wu1273, peng2021nonabelian, bouhon2019nonabelian}, which quantifies the obstruction for pairs of Weyl nodes to leave the $C_2T$ plane. 
For a generic, multi-band system obeying $C_2T$ symmetry, this obstruction is expressed by an additional set of topological invariants, independent of the Chern numbers, which characterize Weyl nodes inside the $k_z=0$ plane. 
These additional invariants are called ``non-Abelian frame charges,'' and we denote them by $q$. 
A pair of nodes with opposite non-Abelian charges have an Euler class of zero, and a pair of nodes with the same non-Abelian charge carry a nonzero Euler class. 
Importantly, if the Euler class is nonzero then the two nodes are not allowed to exit the plane, neither by merging and moving away from $k_z=0$ nor by annihilating with each other.

In our two-band models, these additional invariants $q$ are given simply by the winding number of the vector multiplying the $\sigma$-matrices~\cite{bouhon2020geometric}, and take values $q=\pm 1$.
For the two-node process shown in Fig.~\ref{fig:simple_disprel}(c), the Euler class has several consequences. First, the out-of-plane nodes (blue) must acquire opposite non-Abelian charges ($q=+1$ and $q=-1$) once they enter the plane, reflecting the fact that this pair is allowed to leave the $C_2T$ plane again, through the reversed process. 
Further, the annihilation of the initial in-plane node (red) with one of two nodes that entered the plane tells us that they must have opposite non-Abelian charges. 
Therefore, the non-Abelian charge $q$ of the remaining node must be the same as that of the initial in-plane node at the beginning of the process. 
In other words, even though the Weyl cone flips its chirality, the non-Abelian charge associated to it remains conserved. 
There could be no annihilation if the nodes of equal chirality (blue) were swapped, since the two merging nodes would then have the same value of $q$.

\section{Transport properties} 
The initial and final stages of both chirality flipping processes are the same. 
This means that the observables for large $|\alpha|$ have the same qualitative behavior for the two processes. 
The difference in the two scenarios is prominent close to $\alpha=0$ where the two Weyl nodes outside of the $C_2T$ invariant plane reach the plane. 

\fig{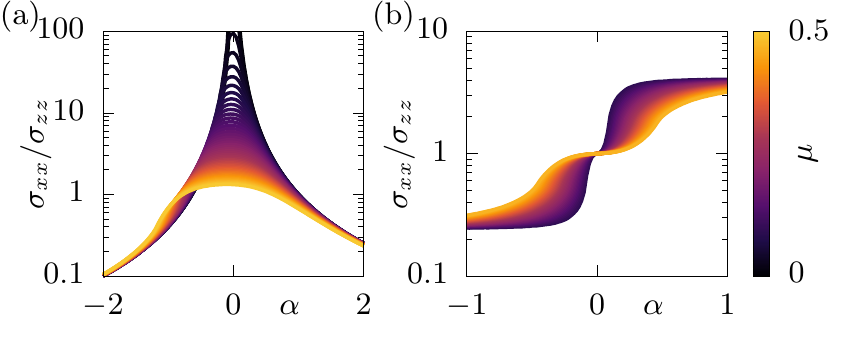}{width=8.6cm}{Ratio of conductivities in the plane of the three Weyl nodes as a function of the parameter $\alpha$ for different chemical potentials $\mu$. Panel (a) shows the three-node process and (b) shows the two-node process.}

We study the differences between the two processes by computing the conductivity using semiclassical Boltzmann transport theory in the constant relaxation time approximation. 
The details of the calculations are shown in the supplemental material (SM) \footnote{See the Supplemental Materials for: $(i)$ details of the conductivity calculation, $(ii)$ details of the \textit{ab initio} results, and $(iii)$ animations of the Weyl-node dynamics as obtained from the model Hamiltonians and from the \textit{ab initio} calculations. Input files used in this study as well as relevant data are available in the repository at \cite{zenodo}}.
At large $|\alpha|$ and low energies, both systems show a similar behavior: the conductivity of isolated Weyl nodes is $\sigma_{jj}\propto\mu^2$ \cite{Tabert2016}, with $j=x,y,z$ and $\mu$ the chemical potential measured relative to the Weyl node energy. 
At $\alpha=0$, however, the two systems are significantly different. 
In the three-node process we have a $\mathcal{C}=-1$ node, with linear dispersion in $k_x$ and $k_y$ and cubic dispersion in the $k_z$ direction. 
In the two-node process we have a double Weyl point with $\mathcal{C}=-2$ and with quadratic dispersion in $k_x$ and $k_z$, but linear dispersion in $k_y$. 
This difference in the dispersion at low energies results in different conductivities as a function of chemical potential.
At $\alpha=0$ the three-node process is characterized by
\begin{align}
    \sigma_{xx}&\propto\mu^{4/3} &\sigma_{yy}&\propto\mu^{4/3} & \sigma_{zz}&\propto\mu^{8/3},
\end{align}
whereas the two-node process shows
\begin{align}
    \sigma_{xx}&\propto\mu^{2} &\sigma_{yy}&\propto\mu & \sigma_{zz}&\propto\mu^{2}.
\end{align}
Note that in the latter case we only focused on the double Weyl node forming at $k_z=0$. The third Weyl node will contribute additively to the conductivity \cite{Note2}.

These differences are prominent when looking at the conductivities in the plane of the three nodes ($\sigma_{xx}$ and $\sigma_{zz}$). 
In Fig.~\ref{fig:cond_main_log} we plot the ratio $\sigma_{xx}/\sigma_{zz}$ as a function of $\alpha$ for different chemical potential values. 
Panel (a) shows the three-node process, in which the ratio is enhanced close to $\mu=0$ and $\alpha=0$, which is due to the different dispersions in the two directions. 
On the contrary, panel (b) shows that $\sigma_{xx}/\sigma_{zz}=1$ at $\alpha=0$ for the two-node process, because of the identical dispersion relations in the two directions.

\section{Weyl nodes in strained M\lowercase{o}T\lowercase{e}$_2$}
 Many materials reported to host Weyl cones have $C_2T$ symmetry, and in several cases Weyl nodes are placed inside the $C_2T$-invariant plane,  including \{W,Mo\}Te$_2$ \cite{PhysRevX.6.031021,PhysRevLett.121.136401}, Ta\{Ir,Rh\}Te$_4$ \cite{PhysRevB.93.201101} and ZrTe \cite{PhysRevB.94.165201}. 
 Under a suitable external perturbation, these compounds are natural candidates for the three-node process described above. 
 As an example, we show based on density-functional calculations that uniaxial strain in MoTe$_2$ controls a process of the type sketched in Fig. \ref{fig:simple_disprel}(b).
 We perform fully relativistic calculations as implemented in the FPLO code \cite{PhysRevB.59.1743}.
 We use as equilibrium lattice parameters those reported in Ref.~\cite{PhysRevX.6.031021} at \SI{100}{\kelvin} ($a=\SI{3.468}{\angstrom}$, $b=\SI{6.31}{\angstrom}$ and $c=\SI{13.861}{\angstrom}$). 
 We focus on uniaxial strain such that the lattice parameter $a$ is enlarged.
 For a fixed deformation $\delta^a$ (measured in percent of the original value of $a$), we determine the deformations of $b$ and $c$ according to the Poisson ratios $\gamma_{ab} = 0.19$ and $\gamma_{ac}=0.96$ reported in Ref.~\cite{yang2017elastic}. Previous works have shown that treating the  Mo-4d shell with the GGA+U method improves the description of photoemission experiments provided by GGA \cite{PhysRevLett.121.136401,huang2019polar,Aryal2019,Singh2020}. 
 Thus we here use the former method with $J = 0$ and $U = \SI{2.6}{\electronvolt}$.
 Further details, including the robustness of our results to all these choices are presented  in the SM \cite{Note2}.

 \begin{figure}[t]
 \centering
  \includegraphics[width=\linewidth]{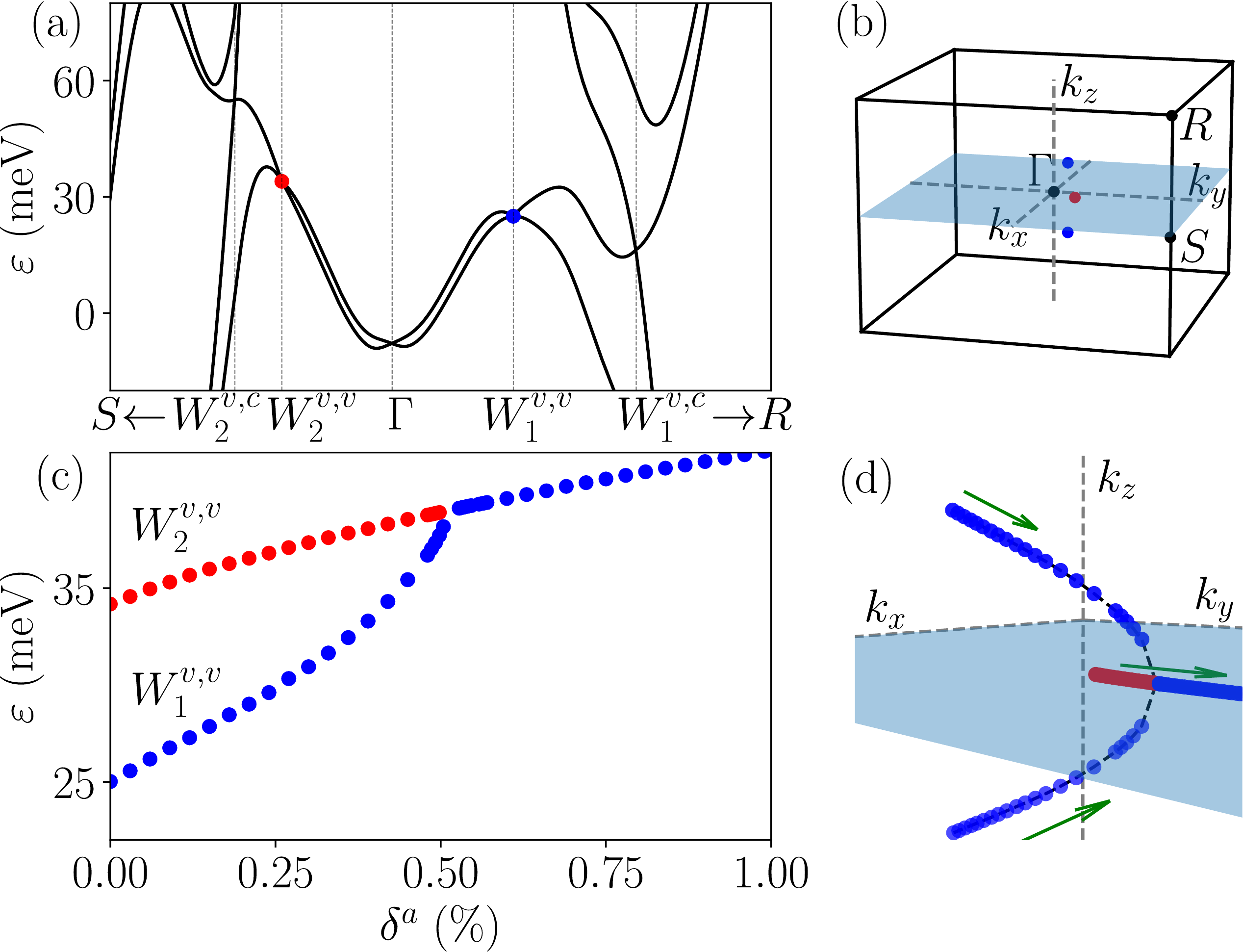}
 \caption{
 (a) Band structure of MoTe$_2$. $W^{v,c}_1$ ($W^{v,c}_2$) is a Weyl node  outside (inside) the $k_z=0$ plane, and connects valence and conducting bands. 
 Analogously, $W^{v,v}_1$  and $W^{v,v}_2$ connect the two upper valence bands. Blue (red) corresponds to negative (positive) chirality.
 (b) Brillouin zone including the Weyl nodes $W^{v,v}_1$ and $W^{v,v}_2$ having positive $k_x, k_y$ coordinates. 
 The $k_z=0$ plane is indicated in blue.
 (c),(d) Energy- and momentum-trajectories of $W^{v,v}_1$ and $W^{v,v}_2$ as a function of tensile strain. 
 Green arrows indicate the trajectories upon increasing the tensile strain. 
 }
 \label{fig:dft} 
 \end{figure}

 Fig. \ref{fig:dft}(a) shows the band structure of MoTe$_2$ without strain ($\delta^a$=0) along a path that includes Weyl nodes connecting the higher valence and lower conducting bands ($W^{v,c}_1$ and $W^{v,c}_2$)  and nodes between the two upper valence bands ($W^{v,v}_1$ and $W^{v,v}_2$). 
 The results of interest in this work originate in the latter two which are found \SI{25}{\milli\electronvolt} ($W^{v,v}_1$) and \SI{34}{\milli\electronvolt} ($W^{v,v}_2$) above the Fermi level. 
 Note that both are type-II, or over-tilted nodes, unlike the band touching points of the toy models presented above.
 $W^{v,v}_1$ has seven other partners in the BZ, which together form a set of eight nodes at generic momentum coordinates, which are related to each other by the reflection symmetries $M_x$ and $M_y$, the twofold rotation around the $c$ axis, $C_{2}$, and time-reversal $T$. 
 On the other hand, there are only four nodes of the type $W^{v,v}_2$, all of which are positioned inside the $C_2 T$-invariant plane [Fig.~\ref{fig:dft}(b)].
 As $\delta^a$ increases, the energy difference between  $W^{v,v}_1$ and $W^{v,v}_2$ decreases and vanishes at the critical point $\delta^a_c\approx\SI{0.5}{\percent}$ [Fig.~\ref{fig:dft}(c)].

Within our numerical precision, at $\delta^a_c$ the Weyl nodes $W^{v,v}_1$ are annihilated while the nodes $W^{v,v}_2$ change their chirality.
 There are a total of four such processes throughout the entire BZ, one at each quadrant of the $(k_x,k_y)$ plane, due to the presence of reflection symmetries. 
 In a given quadrant, before the Weyl merging, one finds a pair of nodes at finite $k_z$ and Chern number ${\cal C}=-1$, and a single node having $k_z=0$ and ${\cal C}=1$. 
 After the merging, only a single Weyl node having ${\cal C}=-1$ and $k_z=0$ exists [Fig. \ref{fig:dft}(d)].
 Thus, MoTe$_2$ realizes a three-node process of the type discussed previously.
 
  Note that none of the nodes are on the twofold rotation axis, so the triple-Weyl merging is not forced to occur, \emph{i.e.}, it is not protected by symmetry.
 Instead, our results suggest that the three-node process is favored energetically, since it involves a smaller change in the local gap between the bands forming the Weyl cones (see SM \cite{Note2}).

\section{Conclusion} 
We have shown that the topological charge associated to Weyl cones can change sign.
This chirality flip involves three nodes, and may occur generically in Weyl semimetal phases.
We discussed the two generic mechanisms by which this can happen: a three-node process where a triplet of Weyl cones merge at the same point in the BZ, and a two-node process, which occurs by means of successive, pairwise mergings of nodes.
Additional symmetries, in our case time-reversal and twofold rotation, increase the likelihood of chirality flips by constraining the positions and relative charges of Weyl nodes. 
In some cases, symmetries may even enforce the occurrence of the three-node process.

Our results indicate that the chirality flip occurs in one of the most well-studied Weyl materials, MoTe$_2$, where moderate uniaxial strain leads to a simultaneous merging of three nodes close to the Fermi level.
Depending on the doping level of different MoTe$_2$ samples, this indicates that chirality flips are within reach of photoemission and transport experiments.
It could be interesting to extend our study of electronic transport to models relevant to MoTe$_2$, in particular to magnetotransport properties sensitive to the chiral anomaly.

More generally, our work sets the stage for further investigating such chirality converting processes in a wide range of materials and scenarios. As we have mentioned, there is an abundance of Weyl materials obeying $C_2T$ symmetry, many of which host Weyl cones pinned to the $C_2T$-invariant plane. Furthermore, similar processes could be also observed in topological metamaterials that host Weyl nodes such as interacting spin systems \cite{Scherubl2019,Frank2020} or multiterminal Josephson junctions \cite{VanHeck2014,Riwar2016}.

An interesting direction for future work could be to examine the behavior of chiral topological metals, which may obey time-reversal and rotation symmetries, but in which inversion symmetry is strongly broken. In these systems, Kramers' theorem guarantees the presence of Weyl points at the time-reversal invariant momenta (TRIM) of the BZ. When TRIM points lie on a rotation axis, we have shown that chirality flips must occur via a three-node process.

\begin{acknowledgments}
We thank Ulrike Nitzsche for technical assistance and J\'{a}nos K. Asb\'{o}th for useful discussions.
This work was supported by the Deutsche Forschungsgemeinschaft~(DFG, German
Research Foundation) under Germany's Excellence Strategy through the
W\"{u}rzburg-Dresden Cluster of Excellence on Complexity and Topology in Quantum
Matter -- \emph{ct.qmat} (EXC 2147, project-id 390858490).
R.-J.~S.~acknowledges 
 funding from the Marie Sk{\l}odowska-Curie programme under EC Grant No. 842901 and the Winton programme as well as Trinity College at the University of Cambridge.
 J.I.F.~acknowledges the support from the Alexander von Humboldt Foundation.
\end{acknowledgments}
\bibliography{references}

\end{document}


\title{Supplemental material -- Chirality flip of Weyl nodes and its manifestation in strained MoTe$_2$}

\author{Viktor K\"{o}nye}
\affiliation{Institute for Theoretical Solid State Physics, IFW Dresden and W\"urzburg-Dresden Cluster of Excellence ct.qmat, Helmholtzstr. 20, 01069 Dresden, Germany}

\author{Adrien Bouhon}
\affiliation{Nordic Institute for Theoretical Physics (NORDITA), Stockholm, Sweden}

\author{Ion Cosma Fulga}
\affiliation{Institute for Theoretical Solid State Physics, IFW Dresden and W\"urzburg-Dresden Cluster of Excellence ct.qmat, Helmholtzstr. 20, 01069 Dresden, Germany}

\author{Robert-Jan Slager}
\affiliation{TCM Group, Cavendish Laboratory, University of Cambridge, J. J. Thomson Avenue, Cambridge CB3 0HE, United
Kingdom}

\author{Jeroen van den Brink}
\affiliation{Institute for Theoretical Solid State Physics, IFW Dresden and W\"urzburg-Dresden Cluster of Excellence ct.qmat, Helmholtzstr. 20, 01069 Dresden, Germany}
\affiliation{Institute  for  Theoretical  Physics,  TU  Dresden,  01069  Dresden,  Germany}

\author{Jorge I. Facio}
\affiliation{Institute for Theoretical Solid State Physics, IFW Dresden and W\"urzburg-Dresden Cluster of Excellence ct.qmat, Helmholtzstr. 20, 01069 Dresden, Germany}

\date{\today}

\begin{abstract}
	In this supplemental material we discuss the details of the results presented in the main paper. We give a detailed explanation on the calculation of the density of states and conductivity. We present the details of the ab-initio calculation. We provide relevant input files and data in the data repository \cite{zenodo}.
\end{abstract}
\maketitle

\setcounter{equation}{0}
\setcounter{figure}{0}
\setcounter{table}{0}
\setcounter{page}{1}
\renewcommand{\theequation}{S\arabic{equation}}
\renewcommand{\thefigure}{S\arabic{figure}}

\section{Density of states and conductivity of the simple models}

Here we give details on the calculation of the density of states (DOS) and conductivity of the simple models introduced in the main text. 
The two models are
\begin{align}
\label{eq:simpleHapp}
    H_1(\vb{k}) &= k_x\sigma_x + \left(\alpha k_z + \beta k_z^3\right)\sigma_y + k_y\sigma_z,\\
\label{eq:simpleHboringapp}
    H_2(\vb{k}) &= k_y\sigma_x + k_x k_z\sigma_y + (k_x^2-k_z^2+\beta k_x^3-\alpha)\sigma_z.
\end{align}
The eigenvalues of these Hamiltonians are
\begin{align}
    E_{1\pm}(\vb{k}) &= \pm\sqrt{k_x^2+k_y^2+\left(\alpha k_z+\beta k_z^3\right)^2},\\
    E_{2\pm}(\vb{k}) &= \pm\sqrt{k_y^2+k_x^2 k_z^2+\left(k_x^2-k_z^2+\beta k_x^3-\alpha\right)^2},
\end{align}
The dispersion relation of these models together with the Berry curvature are shown in Figs~\ref{fig:berry_t} and \ref{fig:berry_d}.

\fig{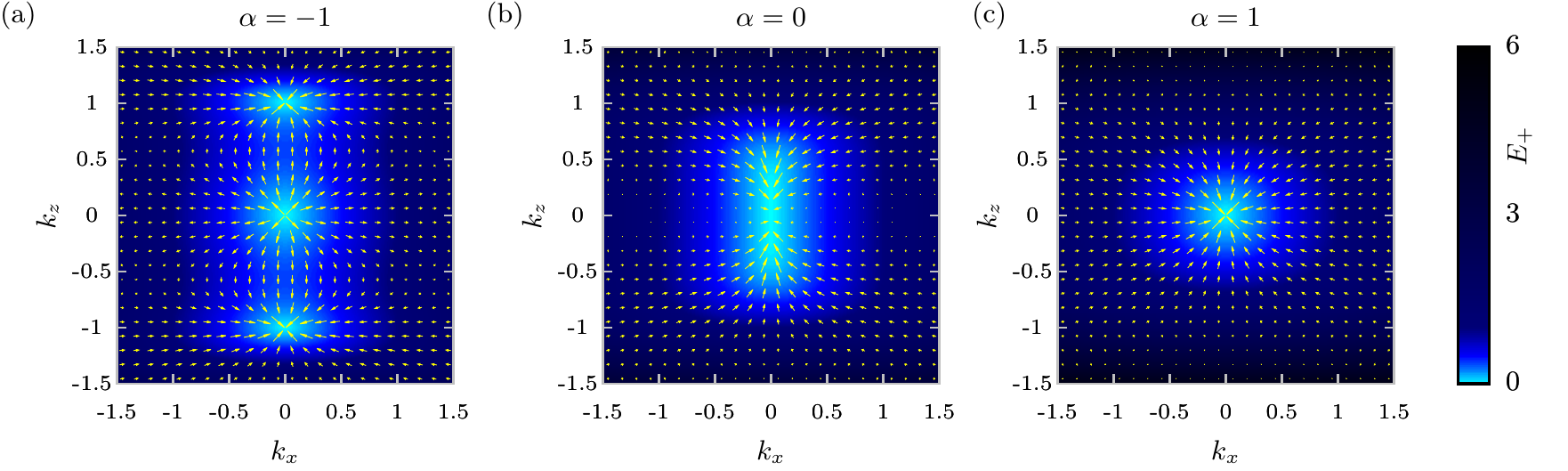}{width=\columnwidth}{Dispersion relation of the $H_1$ Hamiltonian in Eq.~(\ref{eq:simpleHapp}) at three different values of $\alpha$ and $\beta=1$. The yellow arrows show the Berry curvature. $E_+$ denotes the higher-energy band.}
\fig{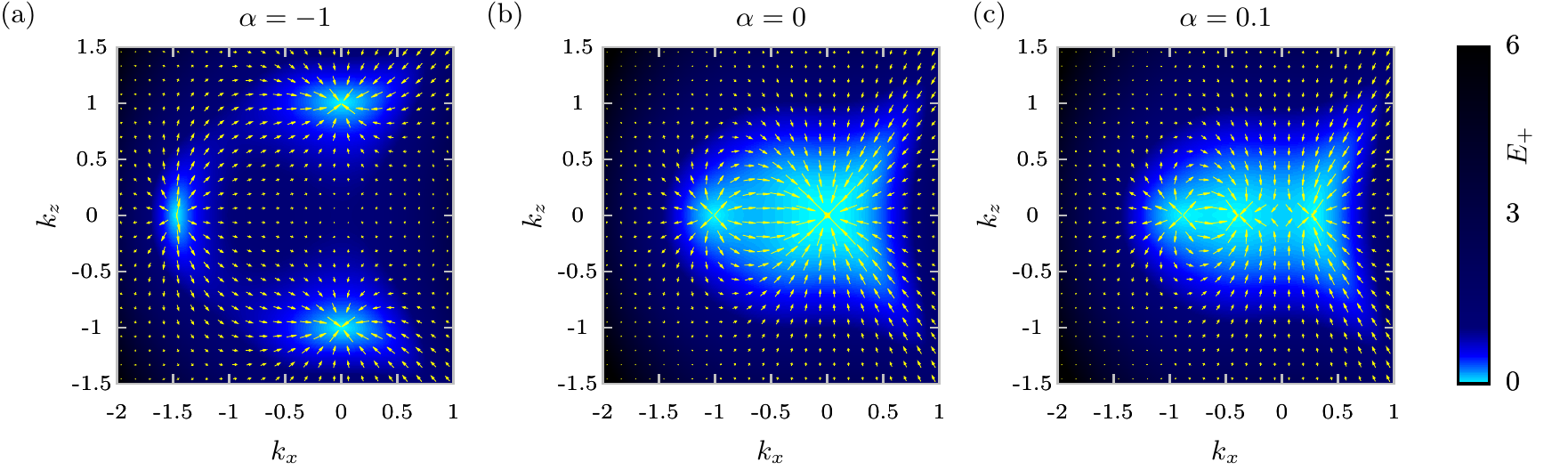}{width=\columnwidth}{Dispersion relation of the $H_2$ Hamiltonian in Eq.~(\ref{eq:simpleHboringapp}) at three different values of $\alpha$ and $\beta=1$. As before, the yellow arrows show the Berry curvature and only the energy of the top band is shown.}

For simplicity we will focus on the part of the processes where the two Weyl nodes at $k_z\neq 0$ enter the $C_2T$-invariant plane. 
In the $H_2$ case this means that we ignore the third Weyl node in the plane and use $\beta=0$.

The density of states is calculated using
\begin{equation}
 D(E) = \sum\limits_\pm \int \frac{\dd[3]{k}}{(2\pi)^3}\delta\left[E_\pm(\vb{k})-E\right],
\end{equation}
where $\delta$ is the Dirac delta function.
Since the spectrum is symmetric with respect to $E=0$, we only discuss $D(E\geq 0)$.

We start with the $H_1$ system. The numerically calculated DOS is shown in Fig.~\ref{fig:dos_t}.

\fig{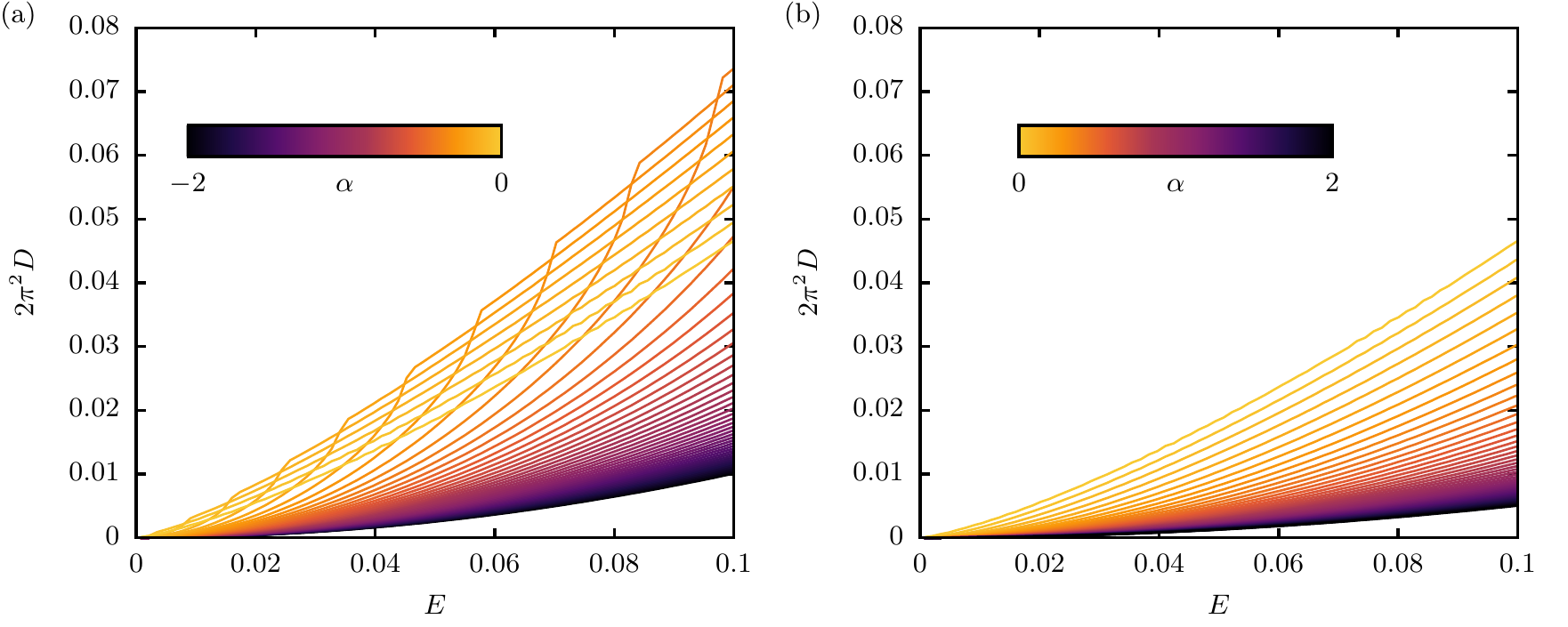}{width=\columnwidth}{Numerically calculated DOS of $H_1$ for different values of $\alpha$ and $\beta=1$. (a) negative values of $\alpha$ with three Weyl nodes. (b) positive values of $\alpha$ with a single Weyl cone.}

In the $\alpha>0$ case at low energies the linear term dominates and we get an anisotropic Weyl node with velocity $\alpha$ in the $z$ direction. In the special case of $\beta=0$ we have linear dispersion up to arbitrary energies and the DOS can be calculated analytically:
\begin{align}
\label{eq:DOSWeyl}
    D_1(E,\alpha,\beta=0) &= \frac{E^2}{2\pi^2\alpha}.
\end{align}
This result matches the curves in Fig.~\ref{fig:dos_t}b when $\alpha>\beta$ and the energy is sufficiently low (the darker curves).

At $\alpha=0$ we are exactly at the chirality-flip point. 
Here in the $z$ direction the linear term vanishes and the cubic term becomes dominant. 
In the case of $\alpha=0$ the DOS can again be calculated analytically and we get
\begin{equation}
    D_1(E,\alpha=0,\beta) = \frac{E^{4/3}}{2\pi^2 \beta^{1/3}}.
\end{equation}
This exact solution matches perfectly the yellow curve in Fig.~\ref{fig:dos_t}b.

At $\alpha<0$ there are three Weyl nodes. 
At low enough energies we see the quadratic increase of the DOS as in Eq.~(\ref{eq:DOSWeyl}). 
Once the energy reaches the saddle points in between Weyl nodes we get a kink in the DOS as seen in Fig.~\ref{fig:dos_t}a. 
The larger the separation between the Weyl nodes the higher the energy where the kink happens. 
Thus, for large negative $\alpha$ the qualitative behavior of the DOS is $D\propto E^2$ in a large energy range. 
Close to $\alpha=0$, however, we get a qualitatively different behavior, with $D\propto E^{4/3}$.

We continue with the $H_2$ system at $\beta=0$. 
The numerically calculated DOS is shown in Fig.~\ref{fig:dos_d}. 

\fig{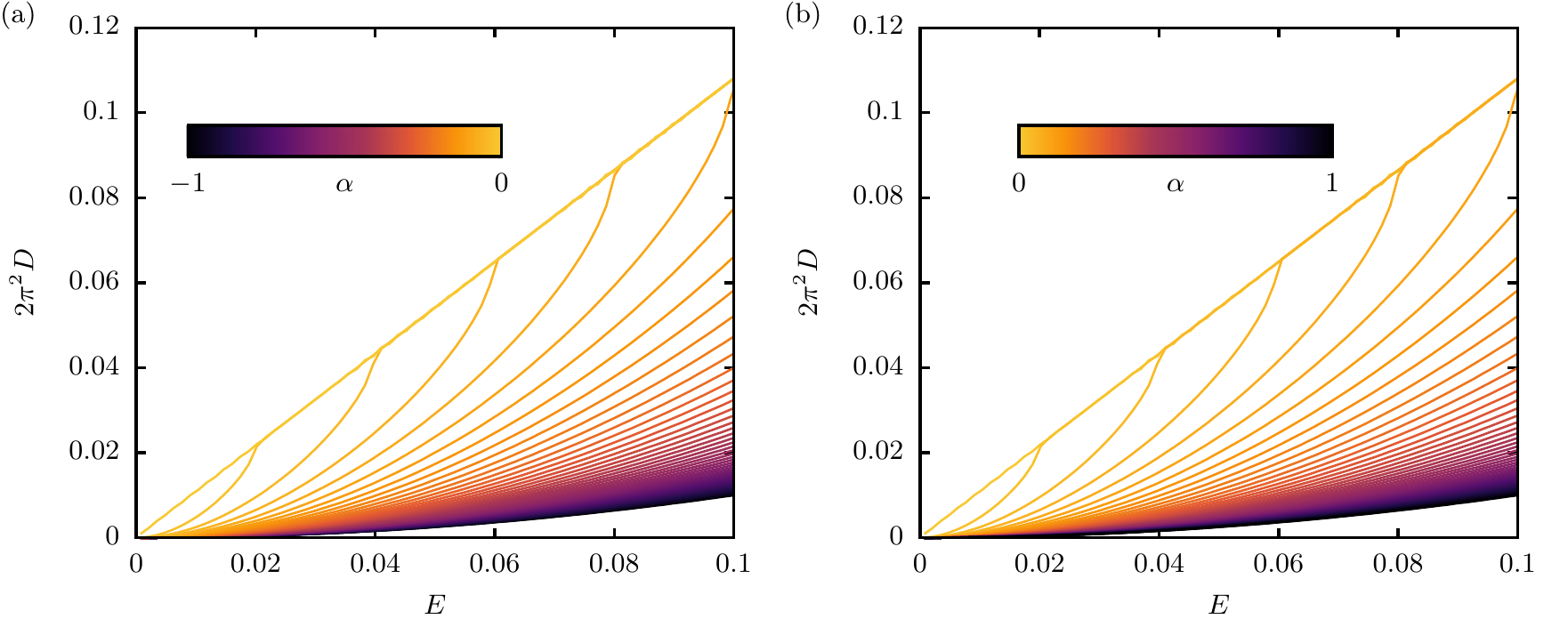}{width=\columnwidth}{Numerically calculated DOS of $H_2$ for different values of $\alpha$ and $\beta=0$. (a) negative values of $\alpha$. (b) positive values of $\alpha$.}

Since the system is symmetric with respect to changing $\alpha\rightarrow -\alpha$ and $k_x\leftrightarrow k_z$ the DOS will be symmetric in $\alpha$.
In the case of $\alpha\neq 0$ at low energies we always have two Weyl nodes and in the DOS we see the quadratic dependency on energy as in Eq.~(\ref{eq:DOSWeyl}). 
Similarly to the $H_1$ system we see kinks in the DOS at the energies of the saddle points in between Weyl nodes. 
The major difference between the DOS of $H_1$ and $H_2$ is at $\alpha = 0$ where the two Weyl nodes outside the $C_2T$ invariant plane reach the plane. 
It can be shown that the DOS for $H_2$ is
\begin{equation}
    D_2(E,\alpha=0,\beta=0) = \gamma E,
\end{equation}
where $\gamma$ is an energy-independent integral

\begin{align}
\label{eq:alpha}
    \gamma &= \frac{2}{\pi^3}\int\limits_0^1\dd{\xi}\int\limits_{-\eta_0}^{\eta_0} \dd{\eta}\frac{\xi}{\sqrt{1-\xi^4\left[2\cosh{(4\eta)}-1\right]}},
    & \eta_0&=\frac{\mathrm{arcosh}{\left(\frac{1}{2\xi^4}+\frac{1}{2}\right)}}{4}.
\end{align}
The numerical value of this integral is $\gamma\approx0.05463$.
As we can see the DOS is proportional to energy, in contrast to the $D \propto E^{4/3}$ dependence we found for the three-node process.

We continue with the calculation of the conductivity of the above Hamiltonians. 
To get a qualitative picture on how the conductivity behaves as a function of the chemical potential we use the semiclassical Boltzmann transport theory with relaxation time approximation \cite{Solyom2009}. 
We also assume zero temperature and use the relaxation time at the Fermi energy as a parameter $\tau\equiv \tau(\mu)$. 
The DC conductivity is calculated as
\begin{equation}
\label{eq:conductivity}
    \sigma_{\nu\nu}(\mu) = e^2\tau \sum\limits_n\int \frac{\dd[3]{k}}{(2\pi)^3}\left[\partial_{k_\nu}E_n(\vb{k})\right]^2\delta(E_n(\vb{k})-\mu),
\end{equation}

The numerically calculated conductivity for the $H_1$ system is shown in Fig.~\ref{fig:cond_t}.

\fig{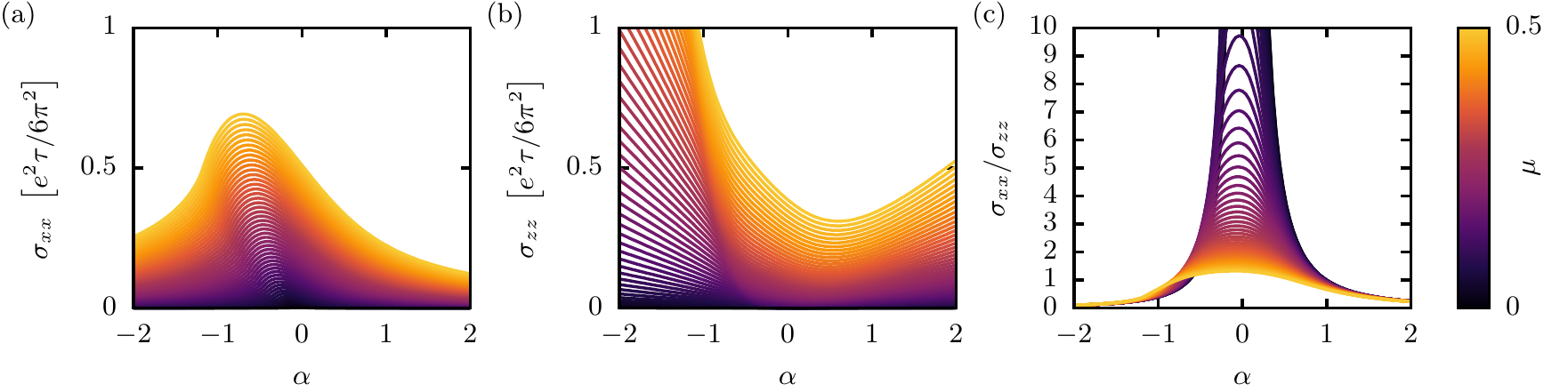}{width=\columnwidth}{Conductivity of $H_1$ calculated from Eq.~(\ref{eq:conductivity}) at different chemical potentials as a function of $\alpha$ for $\beta=1$.}
In the $\alpha=1$, $\beta=0$ case we have a single isotropic Weyl node and for the conductivity we get
\begin{equation}
    \label{eq:condWeyl}
    \sigma_{\nu\nu}(\mu)=\frac{1}{6\pi^2}e^2\tau \mu^2.
\end{equation}
This is the usual result for the conductivity of a single Weyl node \cite{Tabert2016}. 
As it was discussed in Ref.~\cite{Tabert2016}, the Boltzmann approach strictly holds for $\mu>\Gamma$ where $\Gamma = \hbar/2 \tau$ is the scattering rate. 
Close to the Weyl point energy a more detailed description would be necessary, based on the specific type of scattering effects \cite{Burkov2011, Lundgren2014}. 
Since we are interested only in the qualitative behavior we limit ourselves to the simple method of constant relaxation time approximation.

At a general $\alpha$ with $\beta=0$ we have an anisotropic Weyl node and the conductivity can be calculated as
\begin{align}
    \sigma_{xx/yy}(\mu)&=\frac{1}{\alpha 6\pi^2}e^2\tau \mu^2,\\
    \sigma_{zz}(\mu)&=\frac{\alpha}{6\pi^2}e^2\tau \mu^2.
\end{align}
We can see this behavior in Fig.~\ref{fig:cond_t} at $\alpha>0$ and $\alpha\gg \beta$. At the chirality flip point $\alpha=0$ the conductivity is
\begin{align}
    \sigma_{xx/yy}(\mu)&=\frac{3}{14\pi^2}\frac{e^2\tau}{\beta^{1/3}} \mu^{4/3},\\
    \sigma_{zz}(\mu)&=\frac{9}{22\pi^2}e^2\tau \beta^{1/3} \mu^{8/3}.
\end{align}
At $\alpha<0$ we have three Weyl nodes and at small enough chemical potentials the total conductivity is the sum of conductivities of the three nodes. 
The conductivity will be anisotropic but the chemical potential dependence will be quadratic in all directions as in Eq.~(\ref{eq:condWeyl}). 
This can be seen in Fig.~\ref{fig:cond_t}c where the ratio of $\sigma_{xx}$ to $\sigma_{zz}$ is shown. 
Close to $\alpha=\pm1$ the ratio is independent of the chemical potential. 
Close to the chirality flip the anisotropy becomes chemical potential dependent, and the ratio becomes enhanced at low chemical potentials.

We continue with the calculation of the conductivity of the $H_2$ system at $\beta=0$. 
The numerically calculated conductivity for the $H_2$ system is shown in Fig.~\ref{fig:cond_d}.
\fig{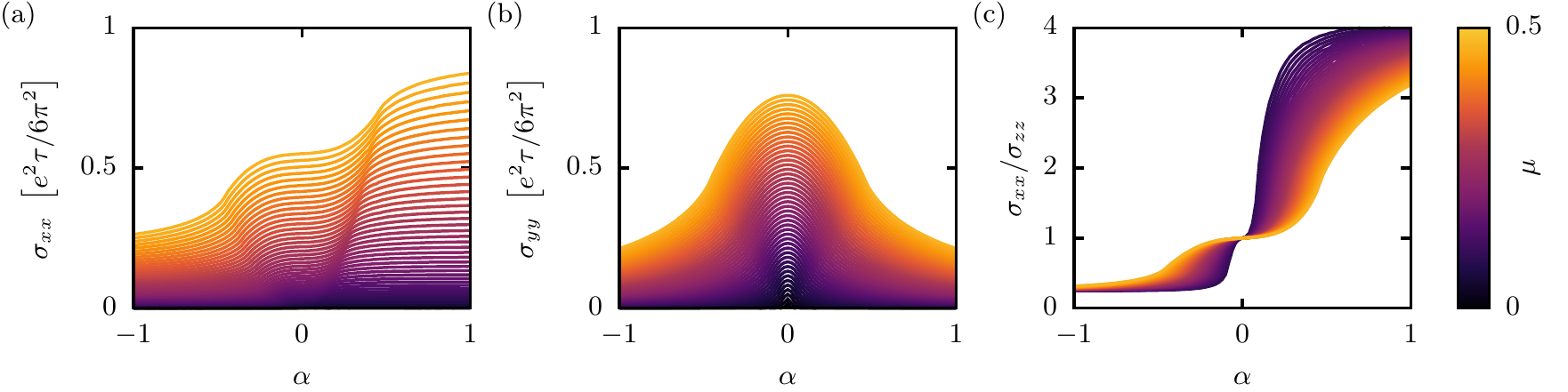}{width=\columnwidth}{Conductivity of $H_2$ calculated from Eq.~(\ref{eq:conductivity}) at different chemical potentials as a function of $\alpha$ for $\beta=0$.}
At $\beta=0$ because of the symmetries of the system $\sigma_{xx}(\alpha) = \sigma_{zz}(-\alpha)$. 
For large $|\alpha|$ we have two Weyl nodes and the conductivity has a quadratic chemical potential dependence as in Eq.~(\ref{eq:condWeyl}). 

At $\alpha=0$, $\sigma_{xx}=\sigma_{zz}$. It can be shown similarly to the DOS that
\begin{align}
    \sigma_{xx/zz}(\mu)&=\lambda_1 e^2\tau \mu^{2},\\
    \sigma_{yy}(\mu)&=\lambda_2 e^2\tau|\mu|,
\end{align}
where $\lambda_1$ and $\lambda_2$ are given by
\begin{align}
    \lambda_1 &= \frac{2}{\pi^3}\int\limits_0^1\dd{\xi}\int\limits_{-\eta_0}^{\eta_0}\dd{\eta} \frac{\xi^7\mathrm{e}^{-2\eta}\left(1-2\mathrm{e}^{4\eta}\right)^2}{\sqrt{1-\xi^4\left[2\cosh{(4\eta)}-1\right]}},\\
    \lambda_2 &= \frac{2}{\pi^3}\int\limits_0^1\dd{\xi}\int\limits_{-\eta_0}^{\eta_0}\dd{\eta} \xi \sqrt{1-\xi^4\left[2\cosh{(4\eta)}-1\right]},
\end{align}
where $\eta_0$ was defined in Eq.~(\ref{eq:alpha}). 
The numerical values of these integrals are $\lambda_1\approx 0.04222$ and $\lambda_2\approx 0.02731$. 

The chemical potential dependence of the conductivity for large $|\alpha|$ is qualitatively the same for the $H_1$ and $H_2$ Hamiltonians, as expected. 
Close to $\alpha=0$, however, the chemical potential difference is qualitatively different, with different power laws in the two cases.

\section{Ab-initio results}

\begin{figure}[h]
\centering
\includegraphics[width=16cm]{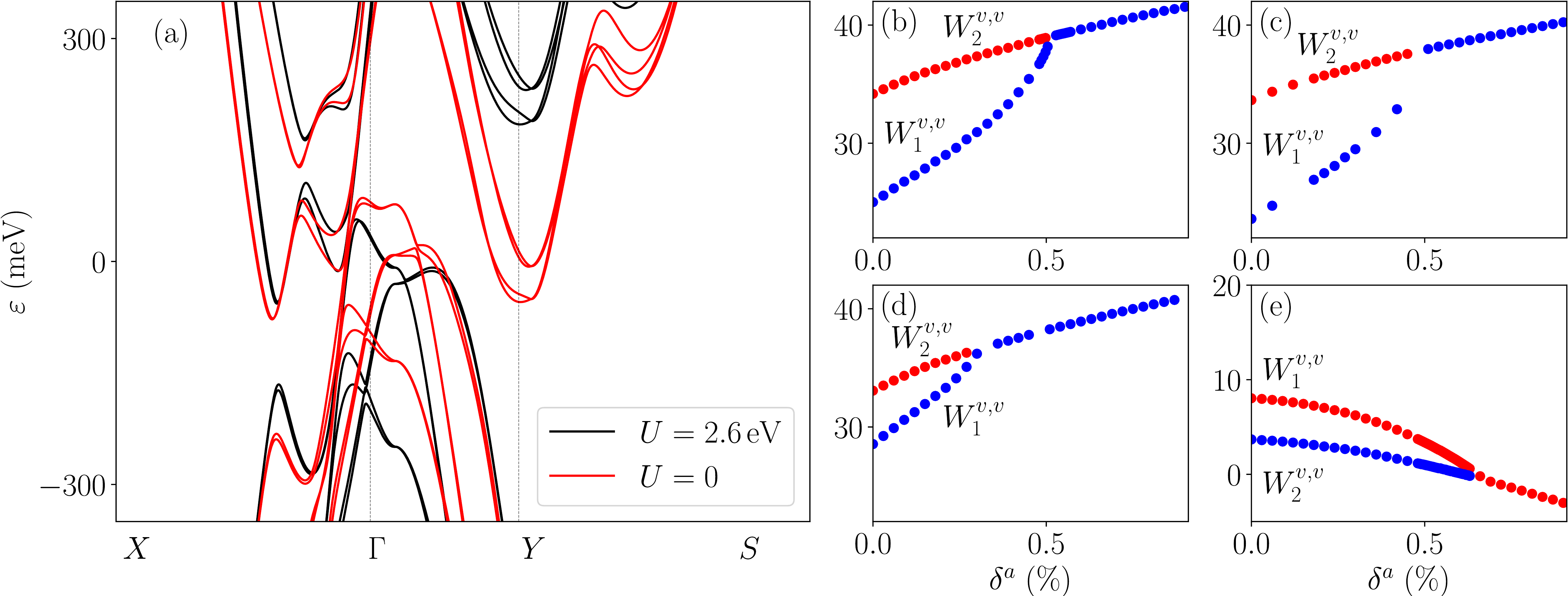} 
\caption{(a) Energy dispersion as obtained with GGA and GGA$+U$ using $U=2.6\,$eV. (b-e) Weyl-node energies as a function of strain as obtained with different calculation schemes: (b)  GGA$+U$ with $U=2.6\,$eV and equilibrium lattice parameters from Ref. \cite{PhysRevX.6.031021}  (as in main text). (c) GGA$+U$ with $U=2.6\,$eV and equilibrium lattice parameters from Ref. \cite{yang2017elastic}. (d) Same as  b but using $U=2.5\,$eV. (e) Same as b but using GGA.
}
\label{fig:bands} 
\end{figure}

We performed fully-relativistic calculations as implemented in the FPLO code version 18.57, using a mesh of $18\times12\times6$ subdivisions for Brillouin zone integrations. Fig.~\ref{fig:bands}a shows the band structures obtained with GGA and with GGA+$U$. 
In the latter case, we use   $J = 0$ and $U = \SI{2.6}{\electronvolt}$. 
In agreement with Ref. \cite{PhysRevLett.121.136401}, the most prominent effect of including $U$ is the elimination of the electron pockets at $Y$. 

Tight-binding Hamiltonians were obtained by constructing Wannier functions associated with the Mo-$4d$ and Te-$5p$ orbitals, using the so-called projecting method. 
The eigenenergies computed with these models typically differ less than 5\,meV from the DFT results. 

Starting from the Wannier model, Weyl nodes were obtained with a bisection-like algorithm that searches for Berry curvature sources, implemented in Ref.~\cite{PhysRevB.93.201101}. 
The algorithm provides a search of candidate Weyl nodes. 
Second, in order to establish or discard Weyl node candidates, we compute  the corresponding Chern number by integrating the Berry curvature around a small sphere centered at the Weyl node candidate position. Table \ref{tab_wp} lists relevant information of the Weyl nodes presented in the main text at zero strain. 
As an additional check, in certain cases we have numerically confirmed the existence of an apparent band-crossing directly in the DFT results by computing the eigenenergies in a sufficiently dense mesh of $k$-points near the Weyl node. 

\begin{table}[h]
	\caption{Weyl nodes for $\delta^a=0$ and $U$$=$\SI{2.6}{\electronvolt}.}
\begin{tabular}{|l|l|l|l|l|}
\hline
Weyl node   & $k_x$ $(2\pi/a)$ & $k_y$ $(2\pi/a)$ & $k_z$ $(2\pi/a)$ & $\varepsilon$ (eV) \\ \hline
$W^{v,c}_1$ & 0.148159         & 0.089422         & 0.038892         & 0.016              \\ \hline
$W^{v,c}_2$ & 0.096388         & 0.029911         & 0                & 0.055              \\ \hline
$W^{v,v}_1$ & 0.073384         & 0.037453         & 0.035651         & 0.025              \\ \hline
$W^{v,v}_2$ & 0.066374         & 0.047740         & 0                & 0.034              \\ \hline
\end{tabular}
\label{tab_wp}
\end{table}

For the calculations of uniaxial strain presented in the main text, we first took steps of $\delta^a$ of $0.03\%$. After that, we made a refinement close to the critical point in steps of $0.006\%$. Very close to the critical point, artifacts of the numerical resolution become evident, e.g., leading to numerically-calculated Chern numbers that deviate significantly from integer values and the corresponding data has been excluded. 

To explore the robustness of the predicted three-node process as a function of strain, we have repeated the search of Weyl nodes by varying the value of $U$ and the values of the equilibrium lattice parameters. 
Fig.~\ref{fig:bands}(b-e) show the results of different tests.  Fig.~\ref{fig:bands}b show data based on GGA$+U$ with $U=2.6$\,eV  and with equilibrium lattice parameters from Ref. \cite{PhysRevX.6.031021} ($a=\SI{3.468}{\angstrom}$, $b=\SI{6.31}{\angstrom}$ and $c=\SI{13.861}{\angstrom}$ ).  Fig.~\ref{fig:bands}c presents data based on the equilibrium lattice parameter reported in \cite{yang2017elastic} ($a=\SI{3.477}{\angstrom}$,$b=\SI{6.335}{\angstrom}$ and $c=\SI{13.883}{\angstrom}$). Variations of this order do not lead to visible changes in the Weyl-node dynamics.
Fig.~\ref{fig:bands}d and e present results based on $U=2.5\,$eV and plain GGA ($U=0$), respectively. The three-body chirality-flip process takes place also for these parameters.  The precise value of $U$ affects the critical value of strain at which the three-body process takes place.

In order to further confirm that the process we observe is indeed the three-node process, we focus on the Weyl node in the $C_2T$-invariant plane. 
If the chirality flip is carried out through two-node processes the annihilation of the $W_2^{v,v}$ node with one of the $W_1^{v,v}$ nodes must happen in the $C_2T$ plane. 
In such a case the velocity difference between the two bands forming the Weyl node in the $C_2T$-invariant plane must vanish at the point of annihilation. 
To exclude this possibility we calculated the minimal, in-plane velocity difference of the bands forming the $W_2^{v,v}$ node. 
For this analysis, we start from two ab-initio constructed Wannier Hamiltonians, one at equilibrium $\delta_a=0$ ($H_0$) and one corresponding to $\delta_a=0.6\%$ ($H_{0.6}$), a value larger than the critical strain at which the three-body process occurs ($\sim 0.5\%$). 
Then, we introduce a parameter $\lambda$ that linearly interpolates between the two: $H[\lambda]=(1-\lambda)H_0+\lambda H_{0.6}$.
The results as a function of $\lambda$ are shown in Fig.~\ref{fig:angle}. 
As we can see no annihilation occurs in the plane, which would be indicated by a dip of the velocity difference. 
Instead, we observe a monotonous increase.

\fig{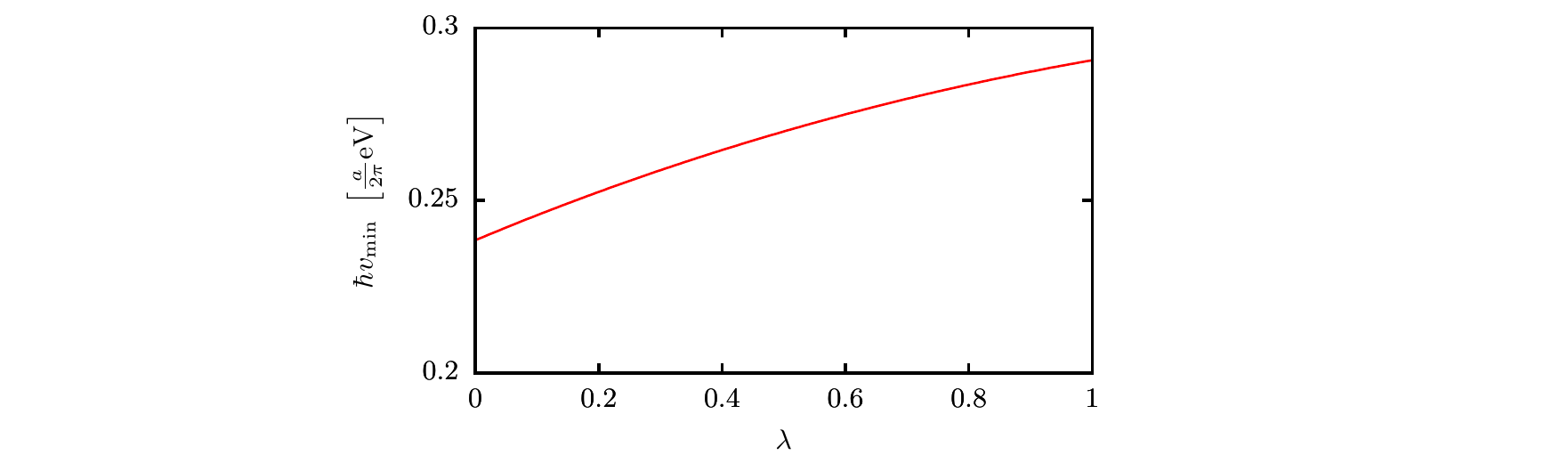}{width=\columnwidth}{Minimal velocity difference of the bands forming the $W_2^{v,v}$ node, computed in the $C_2T$ plane, as a function of strain. $\lambda$ is a parameter that linearly interpolates between the Wannier Hamiltonians at $\delta_a=0$ and $\delta_a=0.6$, values of strain below and above the critical strain ($\sim 0.5\%$). The velocity difference is measured based on the the energy difference on a circle with radius $k=0.001\cdot2\pi/a$ around the Weyl node as $\hbar v=\Delta E/k$.} 

Finally, to gain more understanding on why MoTe$_2$ exhibits a three-node, as opposed to a two-node process, we plot in Fig.~\ref{fig:fermi} the momentum dependent energy gap in the plane containing the three merging Weyl cones. We observe that the smallest energy difference between bands occurs on an arc connecting all three nodes. This suggests that the three-node process (solid arrows) is energetically favored compared to a hypothetical two-node process, for instance as indicated by the dashed arrows. This is due to the fact that having the out-of-plane nodes $W_1^{v,v}$ merge at $k_z=0$ without the in-plane node, $W_2^{v,v}$, would require a larger change in the gap.

\fig{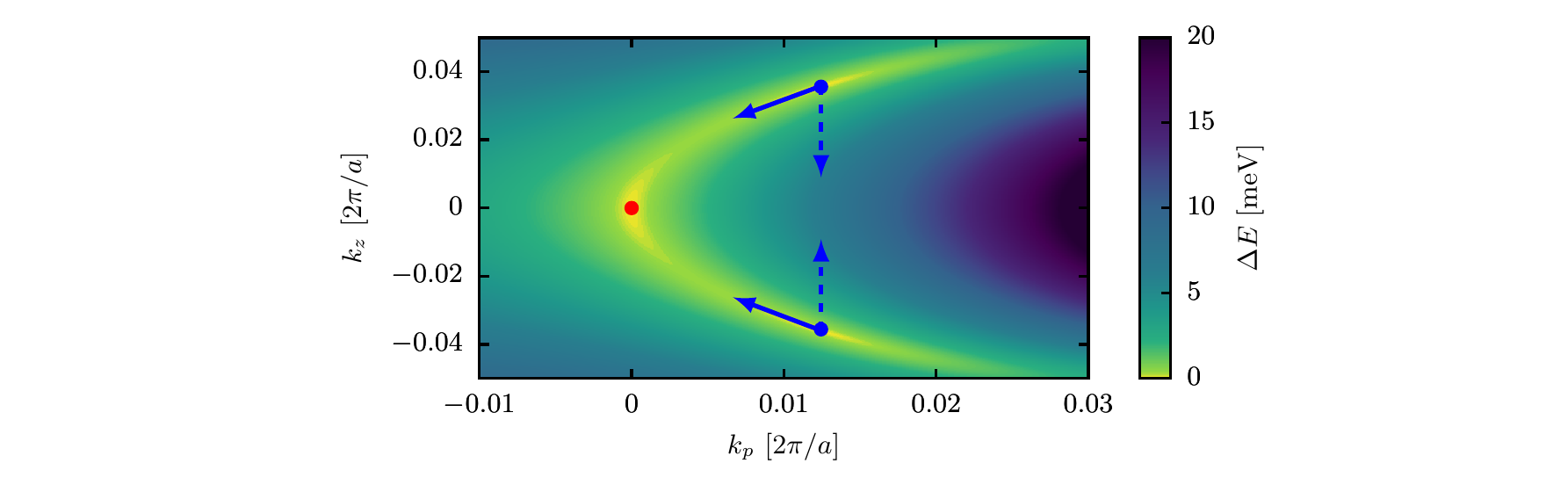}{width=\columnwidth}{Local gap between the two bands that form the Weyl cones $W_1^{v,v}$ and $W_2^{v,v}$. The $k_p$ momentum direction points from the $W_2^{v,v}$ Weyl node (red) to the projection of the $W_1^{v,v}$ nodes (blue) on the $k_z=0$ plane. The solid blue arrows indicate the trajectory taken by the $W_1^{v,v}$ nodes upon increasing strain. The dashed lines indicate a hypothetical path that would achieve the two-node process.}

\bibliographystyle{apsrev4-2}
\bibliography{references}